\documentclass[11pt, letterpaper]{article}

\usepackage{amsmath}
\usepackage{amssymb}
\usepackage{amsthm}
\usepackage{graphicx}
\usepackage{hyperref}
\usepackage{url}
\usepackage{algorithm}
\usepackage{algorithmic}
\usepackage{caption}
\usepackage{subcaption}
\usepackage{booktabs}
\usepackage{authblk}
\usepackage{appendix}
\usepackage[utf8]{inputenc}
\usepackage[T1]{fontenc}
\usepackage[english]{babel}
\usepackage{geometry}
\usepackage{xcolor}
\usepackage{fancyhdr}
\usepackage{microtype}
\usepackage{parskip}
\usepackage{tabularx}
\usepackage{enumitem}
\usepackage{soul}
\usepackage{tabularx}
\usepackage{booktabs} 
\usepackage{listings}

\hypersetup{
    colorlinks=true,
    linkcolor=blue,
    filecolor=magenta,      
    urlcolor=cyan,
    citecolor=red,
}

\definecolor{headercolor}{RGB}{0, 50, 100}
\title{Benchmarking a foundation LLM on its ability to re-label structure names in accordance with the AAPM TG-263 report}

\author[1]{Jason Holmes, PhD\thanks{corresponding author: holmes.jason@mayo.edu}}
\author[1]{Lian Zhang, PhD}
\author[1]{Yuzhen Ding, PhD}
\author[1]{Hongying Feng, PhD}
\author[2]{Zhengliang Liu, MS}
\author[2]{Tianming Liu, PhD}
\author[1]{William W. Wong, MD}
\author[1]{Sujay A. Vora, MD}
\author[1]{Jonathan B. Ashman, MD, PhD}
\author[1]{Wei Liu, PhD}

\affil[1]{Department of Radiation Oncology, Mayo Clinic, Phoenix, Arizona, 85054}
\affil[2]{School of Computing, University of Georgia, Athens, Georgia, 30602}

\date{}

\begin{document}

\maketitle

\begin{abstract}

\noindent\textbf{Purpose:}
To introduce the concept of using large language models (LLMs) to re-label structure names in accordance with the American Association of Physicists in Medicine (AAPM) Task Group (TG)-263 standard, and to establish a benchmark for future studies to reference.

\noindent\textbf{Methods and Materials:}
The Generative Pre-trained Transformer (GPT)-4 application programming interface (API) was implemented as a Digital Imaging and Communications in Medicine (DICOM) storage server, which upon receiving a structure set DICOM file, prompts GPT-4 to re-label the structure names of both target volumes and normal tissues according to the AAPM TG-263. Three disease sites, prostate, head and neck, and thorax were selected for evaluation. For each disease site category, 150 patients were randomly selected for manually tuning the instructions prompt (in batches of 50) and 50 patients were randomly selected for evaluation. Structure names that were considered were those that were most likely to be relevant for studies utilizing structure contours for many patients.

\noindent\textbf{Results:}
The overall re-labeling accuracy of both target volumes and normal tissues for prostate, head and neck, and thorax cases was 96.0\%, 98.5\%, and 96.9\% respectively. Re-labeling of target volumes was less accurate on average except for prostate - 100\%, 93.1\%, and 91.1\% respectively.

\noindent\textbf{Conclusions:}
Given the accuracy of GPT-4 in re-labeling structure names of both target volumes and normal tissues as presented in this work, LLMs are poised to be the preferred method for standardizing structure names in radiation oncology, especially considering the rapid advancements in LLM capabilities that are likely to continue.

\end{abstract}

\section{Introduction}
A long-standing problem in radiation oncology is inconsistent structure name labeling of both target volumes and normal tissues \cite{cancers15030564, SCHULER2019191, SLEEMANIV2020103527, https://doi.org/10.1002/acm2.13662, healthcare8020120, biomedinformatics3030034, https://doi.org/10.1111/1754-9485.12892, Miller_2017, CAISSIE2023100925, NYHOLM2016344, kairn2016bulk, lee2022radiotherapy, https://doi.org/10.1002/acm2.12701, zheng2021nomenclature}. Structure contours serve as a foundation for treatment planning, dosimetric evaluations, and patient outcomes studies. With a growing emphasis on evidence-based medicine and the constant influx of new technologies, such as machine learning\cite{RN1255,yang2022empirical,yang2021seed, yang2021exploratory, nathan2022cardiopulmonary, nathan2020early, bhangoo2020acute, bhangoo2021intensity}, artificial intelligence (AI)\cite{RN1254,zhang2023beam, ding2022accurate, balagopal2021psa, rezayi2022clinicalradiobert, liao2023mask}, and more recently artificial generative intelligence (AGI)\cite{10.3389/fonc.2023.1219326, liu2023radonc, zhang2023segment, liu2023deid, dai2023chataug, wu2023exploring}, the importance of accurate and consistent structure name labeling is increasing\cite{MAYO2023533}. The issue of inconsistent structure name labeling is not a sporadic challenge; it is a persistent hurdle faced regularly by practitioners, researchers, and technologists in the field.

To address this problem, the American Association of Physicists in Medicine (AAPM) created a document known as the AAPM Task Group (TG)-263\cite{MAYO20181057} in 2018, which defined a standard for naming structures including both target volumes and normal tissues. Yet, despite having a defined standard, the complexity of the AAPM TG-263 report has made its implementation in clinical practice challenging, resulting in many clinics still not adopting it. Even for clinics that have adopted the AAPM TG-263 standard, data prior to its adoption likely remains unchanged, creating a barrier for large-scale data mining studies. For this reason, researchers have investigated methods for re-labeling structure names, both prospectively and retrospectively. 

Previous re-labeling methods have typically employed the use of look-up tables \cite{SCHULER2019191}, templates \cite{https://doi.org/10.1002/acm2.12701}, or machine learning \cite{cancers15030564, SLEEMANIV2020103527, https://doi.org/10.1002/acm2.13662, healthcare8020120, biomedinformatics3030034}. The inputs for machine learning approaches have typically included the structure names and structures converted to binary masks. Inputs may also include CT information, dose information, or the structure volume sizes. The inclusion of additional information beyond just the structure names has improved results. However, these studies have consistently highlighted a crucial finding: the structure name itself is the most significant factor in the accuracy of these models to correctly re-label structure names.

While prior methods have been relatively successful in distinguishing and organizing structure names, albeit typically for a small selection of structures, none were both fully automated and universal (able to re-label any structure name). Furthermore, these methods do not directly interpret AAPM TG-263 guidelines. Being that the AAPM TG-263 is a textual report, large language models (LLMs) are a natural choice for interpreting and following the guidelines contained in the AAPM TG-263 as they can be provided as direct input to LLMs. Taking into account the significance of the structure name text in prior studies, we introduce the concept of using foundation LLMs to standardize structure names, providing the AAPM TG-263 guidelines as direct input alongside the structure names. Since Generative Pre-trained Transformer (GPT)-4 (OpenAI, San Francisco, CA)\cite{openai2023gpt4} has shown to be quite knowledgeable on the topic of radiation oncology\cite{huang2023benchmarking, holmes2023evaluating, liu2023radonc, liu2023artificial, waters2023unlocking}, we benchmark the performance of GPT-4 in correctly re-labeling structure names in accordance with the AAPM TG-263.

\section{Methods and Materials}
\label{sec:methods}
For evaluating the accuracy of GPT-4, defined as the percentage of structure names that were correct after re-labeling, three disease sites were considered: prostate, head and neck, and thorax. For each disease site, 150 patients were selected for manually tuning the instructions prompt (in batches of 50) and 50 patients were selected for evaluation - a total of 600 patients. All patient data utilized in this study were selected at random from the patient database of Aria (Varian Medical Systems, Palo Alto, CA) at the Department of Radiation Oncology of Mayo Clinic in Arizona, spanning between the years 2014 and 2023. Only structure names (thus no patient information) were provided to GPT-4.

A difficult aspect of this study was to decide which structure names should be considered in determining the re-labeling accuracy of GPT-4. Most studies consider only 5-10 of the most important structures for a particular disease site. Taking into account the generalization abilities of LLMs, we wanted to consider as many structure names as possible, albeit while still allowing for a reasonable comparison of the re-labeling accuracy across different studies. Taking this into consideration, we did not consider structures that did not represent target volumes or normal tissues, for example "dental", "fiducial", etc., or were designated in some unique or non-standard manner, for example "physics", "preop", etc. Furthermore, we did not consider any structure names containing digits, which were usually generated as supplemental structures by dosimetrists for treatment planning or for patient alignment during beam delivery used by therapists per our institutional protocol. While structure names of these kinds can be made compliant with the AAPM TG-263 in many cases, these structure names were not considered for the purpose of evaluating accuracy as they would likely not be used in data-centered studies in radiation oncology. Nevertheless, GPT-4 was instructed to re-label all structure names except those starting with a digit, "x", "z", or those containing "dose". In order to illustrate the effectiveness of GPT-4 in re-labeling structure names as a whole, we also include one example output per disease site (see Appendix).

\subsection{Implementation}
\begin{figure}
    \centering
    \includegraphics[width=1.0\textwidth]{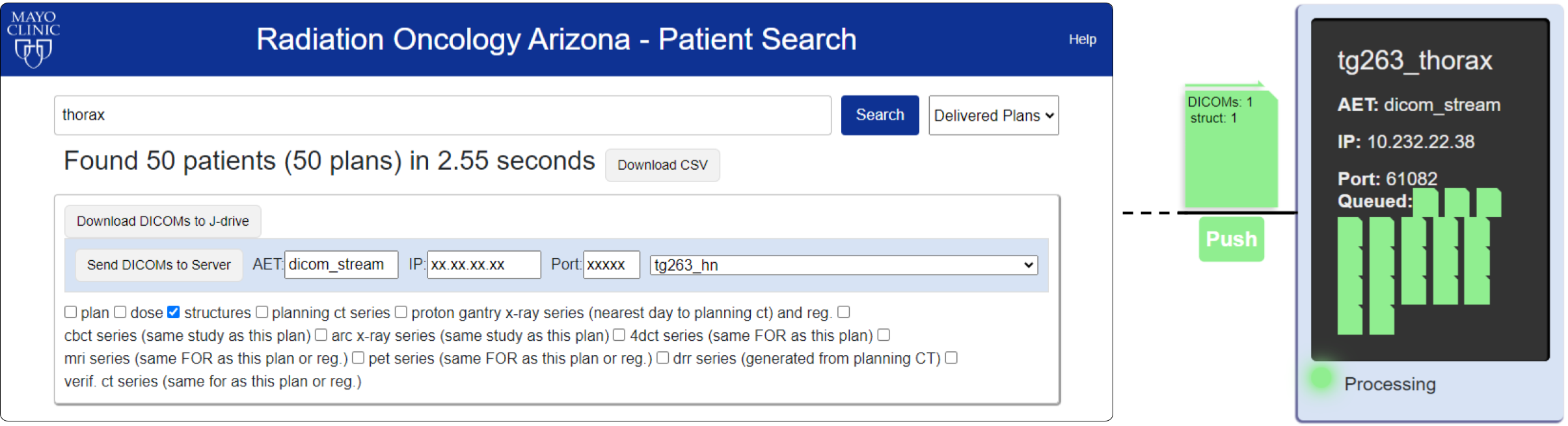}
    \caption{Implementation.}
    \label{fig:implementation}
\end{figure}

The AAPM TG-263 re-labeling software was implemented as a Digital Imaging and Communications in Medicine (DICOM) server with a web monitoring system, which upon receiving a structure set DICOM file, prompts GPT-4, receives a response from GPT-4, then re-labels the original structure names with the AAPM TG-263-standard names. In order to limit the length of prompts, each disease site had its own respective re-labeling application. Our in-house patient search engine was used to select the patients by disease site. Once selected, the patient search website was used to send the structure set DICOM files of each patient to the re-labeling application. See Figure \ref{fig:implementation} for a schematic diagram of the implementation. The web monitoring system allows for downloading of logs generated during the re-labeling process, which can be used for evaluation in clinical research scenarios.

\subsection{Prompts}
Two established prompting techniques were used in this work: zero-shot learning\cite{NEURIPS2022_8bb0d291} and chain-of-thought prompting\cite{NEURIPS2022_9d560961, holmes2023evaluating}. Zero-shot learning is an attempt to teach the model the knowledge required to perform a task for which it was never trained. Chain-of-thought is a strategy where the LLM is prompted to explain its reasoning prior to giving the answer as an attempt to give the LLM a "scratch-pad" from which to work.

The prompts provided to GPT-4 contained six parts:
\begin{enumerate}
    \item The AAPM TG-263 guiding principles for non-target nomenclature.
    \item The AAPM TG-263 guiding principles for target nomenclature.
    \item All the standard structure names provided in the "Nomenclature Worksheet" (part of the AAPM TG-263 effort) corresponding with the disease site.
    \item Institution-specific instructions corresponding with the disease site.
    \item Instructions telling GPT-4 to re-label the structure names in accordance with the AAPM TG-263 and to explain its reasoning prior to answering for each structure name.
    \item The structure names to be re-labeled.
\end{enumerate}

\begin{figure}
    \centering
    \includegraphics[width=0.6\textwidth]{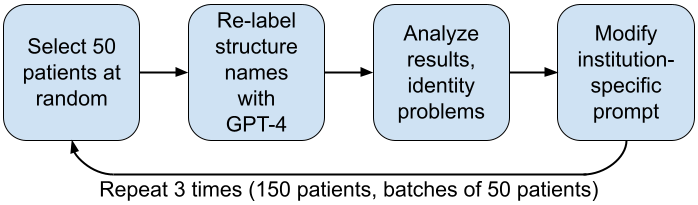}
    \caption{The workflow used to manually tune the institution-specific part of the prompt.}
    \label{fig:prompt}
\end{figure}

There may be very common structure names specific to an institution that are best handled by explicitly defining the structure names within the institution-specific instructions. For example, "BODY should be Body" or "The brain CTV should be CTV\_Brain". In addition, there may be institution-specific structure names that are not defined in the AAPM TG-263 report, such as "STV", "bolus helmet", "couch", etc., that should be defined in the institution-specific instructions. In general, the institution-specific instructions should be carefully crafted to improve accuracy while meeting the clinical needs of the institution. The workflow for tuning the institution-specific prompts used in this study is provided in Figure \ref{fig:prompt}. 

After tuning the institution-specific prompts, 50 randomly selected patients per disease site were selected for evaluation. In analyzing the relabeling accuracy for each disease site, structures were grouped by sub-disease sites. For example, structure names containing "pharyn", "cricopharyngeus", "crico\_p", or "constrict" were classified as being a part of the sub-disease site of "Pharynx".

\section{Results}

The number of structure names considered for each disease site (50 patients per disease site) reflected their overall complexity. For the prostate, there were 207 unique structure names, out of which 86 were considered for determining accuracy. For the head and neck, the numbers were the highest, with 621 unique structure names and 236 being considered for determining accuracy. The thorax disease site had 337 unique names with 183 being considered.

\begin{figure}
    \centering
    \includegraphics[width=0.55\textwidth]{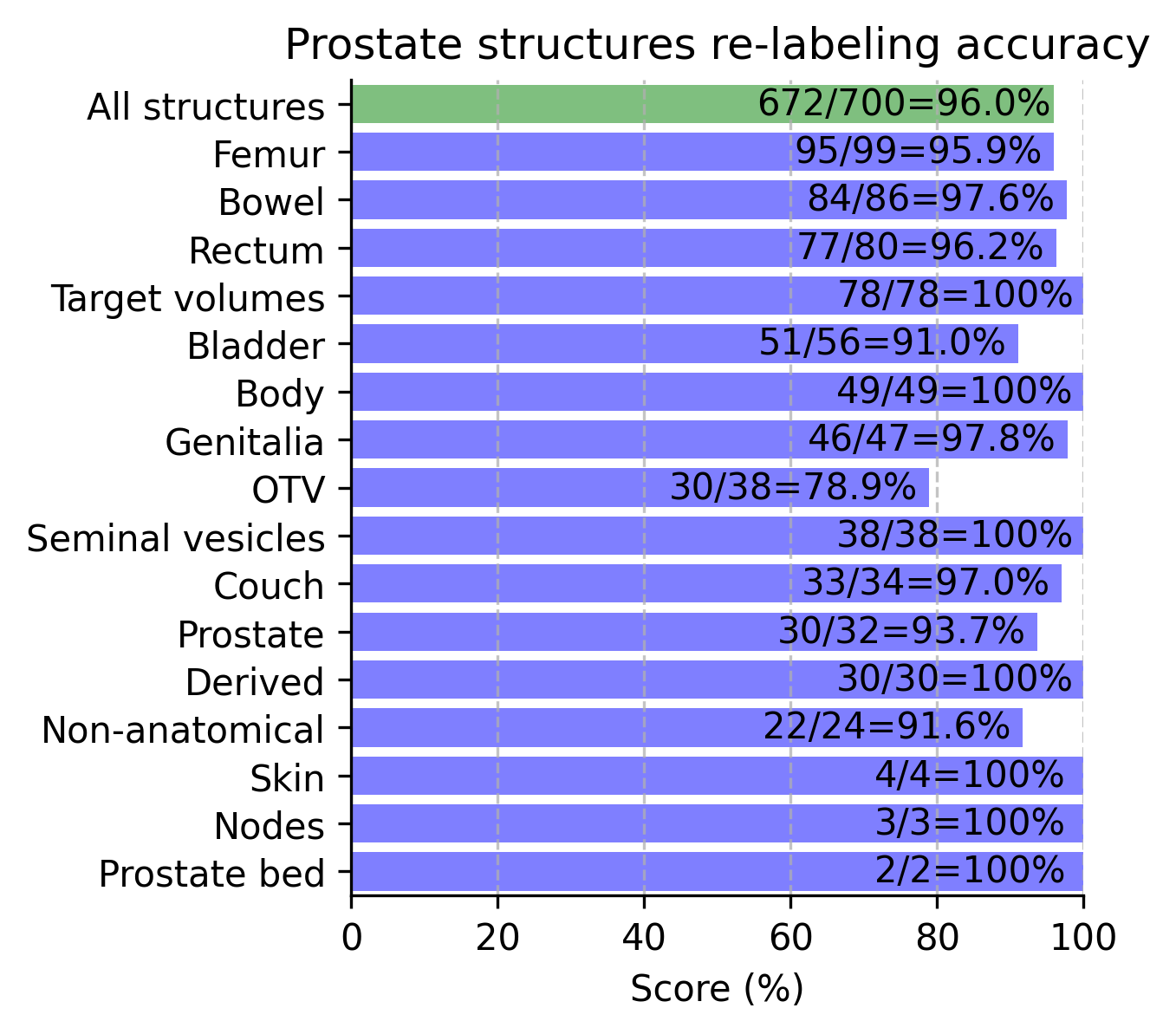}
    \caption{The accuracy of GPT-4 in re-labeling prostate structure names in accordance with the TG-263 report.}
    \label{fig:results_prostate}
\end{figure}

\begin{figure}
    \centering
    \includegraphics[width=0.55\textwidth]{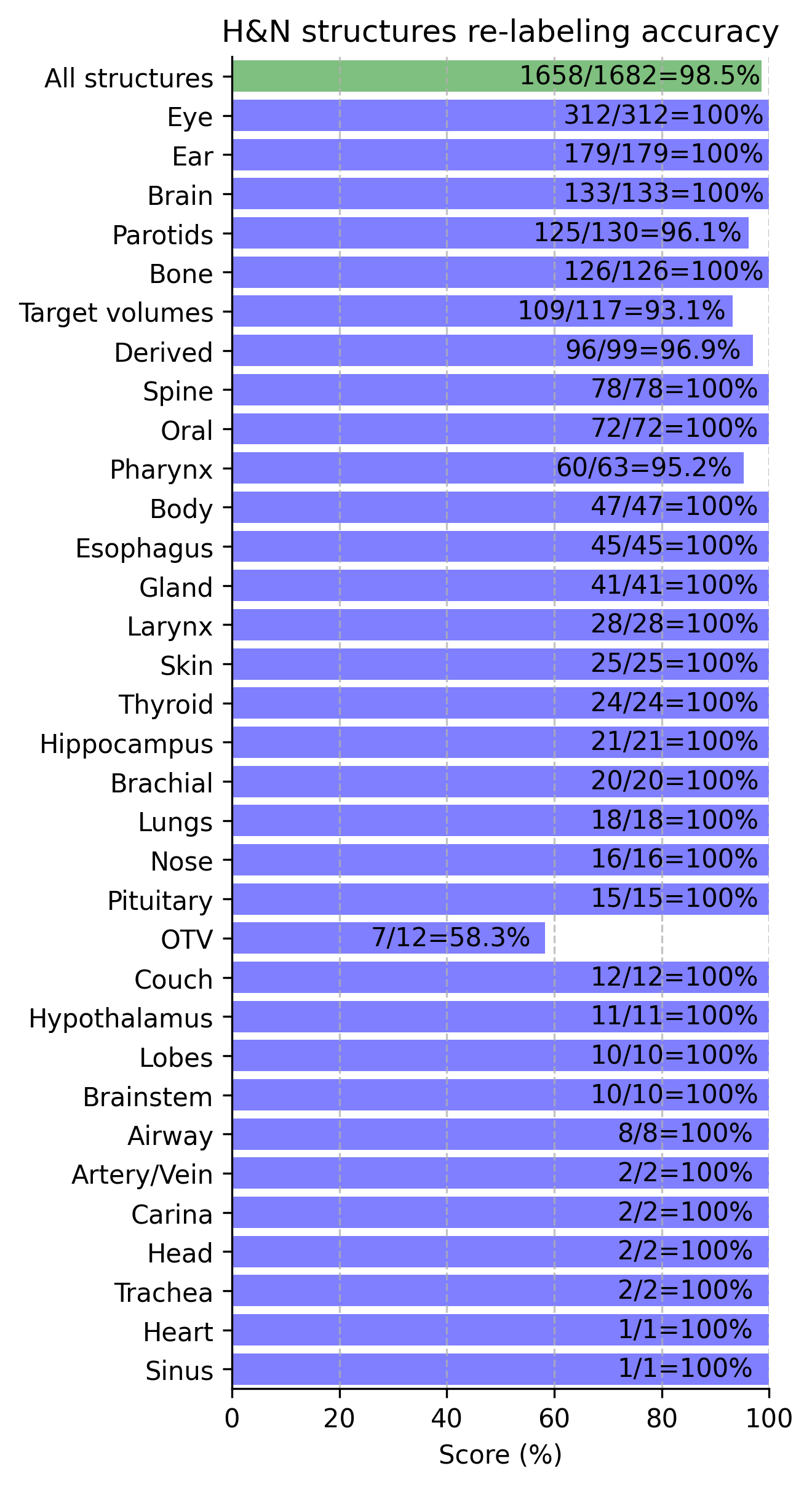}
    \caption{The accuracy of GPT-4 in re-labeling head and neck structure names in accordance with the TG-263 report.}
    \label{fig:results_hn}
\end{figure}

\begin{figure}
    \centering
    \includegraphics[width=0.55\textwidth]{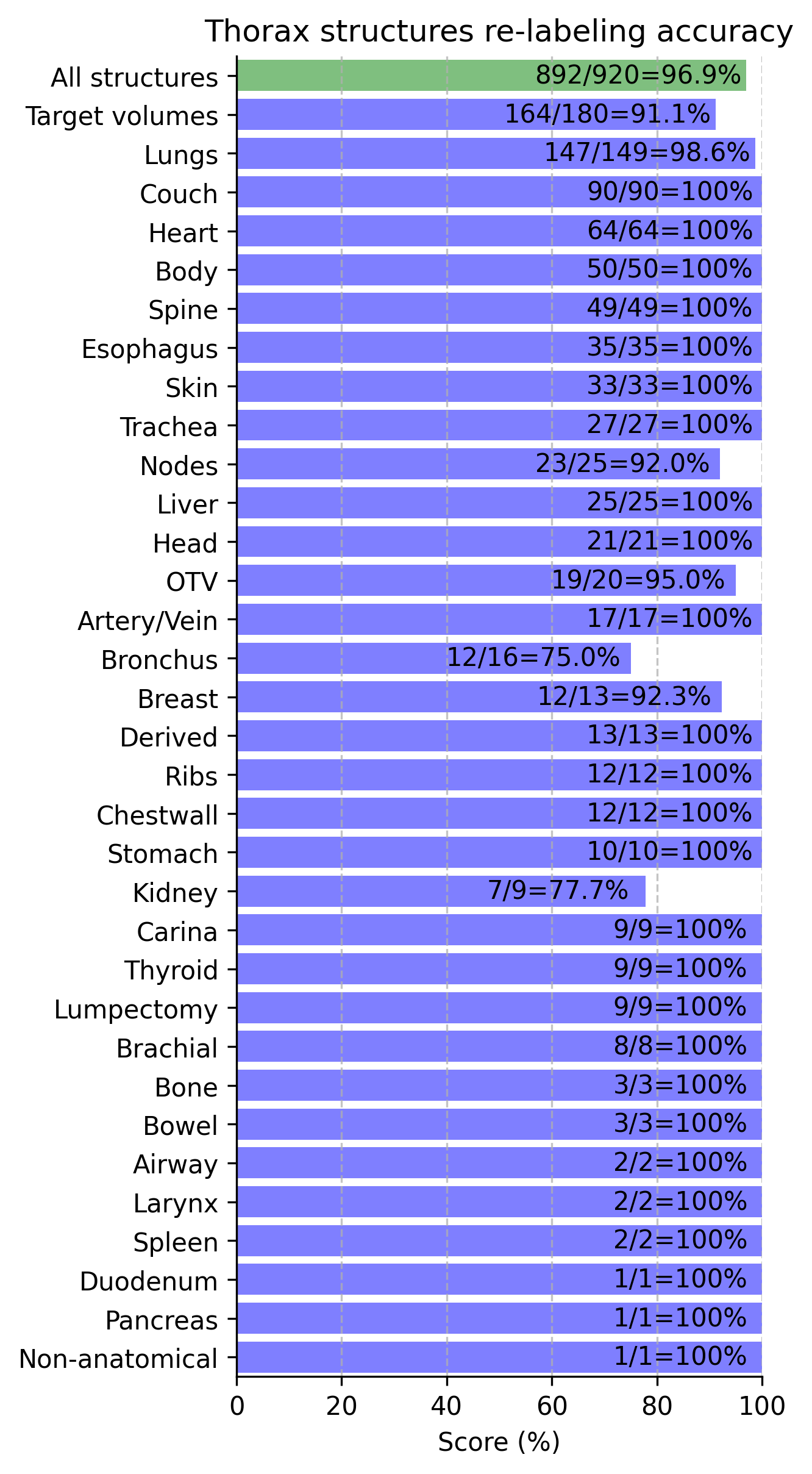}
    \caption{The accuracy of GPT-4 in re-labeling thorax structure names in accordance with the TG-263 report.}
    \label{fig:results_thorax}
\end{figure}

The main results for each disease site are shown in Table \ref{tab:accuracy}. The strcuture name re-labeling accuracy results broken down by sub-disease sites for prostate, head and neck, and thorax are given in Figures \ref{fig:results_prostate}, \ref{fig:results_hn}, and \ref{fig:results_thorax} respectively. The optimization target volume (OTV), which is only used at our institution for proton therapy only, was treated separately from the other target volumes since the performance of GPT-4 varied greatly for OTVs and since OTV is not a standard target volume. The overall accuracy for all disease sites was above 96\%, however the target volumes for head and neck cases and thorax cases dropped to the low 90s.

Most mistakes were typically trivial, such as "fem\_head\_l" being labeled as "Fem\_Head\_L" when it should have been "Femur\_Head\_L". In some cases, the ordering of substrings was incorrect, such as "Lumpectomy\_PTV" instead of "PTV\_Lumpectomy". In some cases, letters were not capitalized when they should have been. 

Out of all the structure names that were re-labeled, there were only 14 mis-classifications, 7 were unique, out of a total of 3,302 structure names that were provided to GPT-4 for re-labeling. For prostate: "OTVProstate" was labeled as "zProstate" on 3 separate occasions. For head and neck: 2 occurrences where the "brain-ptv", a subtraction volume, was labeled as the brain PTV, and 4 occurrences where a target volume was mis-classified as a subtraction volume, for example, "OtvBrain" being labeled as "Brain-OTV". For thorax: 4 occurrences where "bronch\_tree\_prox" was labeled as "AirWay\_Prox" when it should have been labeled as "Bronchus\_Prox".

\begin{table}[h!]
    \centering
    \caption{Overall accuracy per disease site.}
    \begin{tabular}{|c|c|c|}
        \hline
        \textbf{Category} & \textbf{Overall Accuracy} & \textbf{Target Volumes (not OTV)} \\
        \hline
        Prostate & 96.0\% & 100\% \\
        \hline
        Head and Neck & 98.5\% & 93.1\% \\
        \hline
        Thorax & 96.9\% & 91.1\% \\
        \hline
    \end{tabular}
    \label{tab:accuracy}
\end{table}

\section{Discussion}

This work, which used an LLM (GPT-4) to re-label structure names of both target volumes and normal tissues according to the AAPM TG-263, represents one of the first clinical applications of LLMs in radiation oncology with real-world clinical settings. Overall, GPT-4 did an exceptional job at handling highly heterogeneous sets of structure names for each disease site, demonstrating its ability to work in real-world clinical settings. Indeed, this application is being used by our clinic in structure comparison studies as well as for data pre-processing for training AI models that require structure contours.

There are several distinct advantages that foundation LLMs have over alternative AI models in re-labeling structure names according to the AAPM TG-263. LLMs are provided with all the structure names simultaneously and may therefore consider all the provided structure names in determining each new structure name, conceptually very similar to semantic context. Being that the AAPM TG-263 is a text document, LLMs allow for accepting text information from the AAPM TG-263 as direct input. In this work, since GPT-4 is currently limited to 8,000 tokens, we provided only the guiding principles for non-target and target nomenclature as well as the standard structure names given in the "Nomenclature Worksheet" on a per-disease basis. However, in the near future there will be a 32,000 token version of GPT-4 available that would allow for including more of the AAPM TG-263 report, if desired. As LLMs continue to improve, there may come a time when the entire AAPM TG-263 report can be provided as input, allowing for generalization to all disease sites. Finally, as foundation LLMs are updated, for example if/when GPT-5 is released, the code can be updated by simply changing the model from GPT-4 to GPT-5.

Although we did not directly track costs (not possible using GPT-4 API), we attempted to isolate our usage such that we could give an estimate of the cost. Costs were approximately \$0.15 per prostate case, \$0.20 per thorax case, and \$0.25 per head and neck case. While costs are likely to reduce over time per case, costs are proportional to the amount of information provided to the LLM and the number of structure names to be re-labeled, which may increase over time, if for example, the entire AAPM TG-263 report is included.

The poor performance for OTVs may be the result of a possible ambiguity in the AAPM TG-263 report, where the non-target guidelines indicate that optimization volumes should have a "z" placed before the name with an example given as "zPTVopt". This is an odd choice as an example since it combines a target volume with "opt" and is given in the non-target set of guidelines. Should "zPTVopt" be considered a target volume or an optimization volume? In the case of OTV, it would seem that "z" should be used, however there is no "opt" prefix or suffix and so there is a bit of ambiguity that may have led to poor performance for OTVs.

Some anecdotal observations of GPT-4 were as follows: in structure names that were not considered for accuracy, mis-classifications seemed to occur most often for highly unique or ambiguous structure names. The structure names for some patients caused GPT-4 to diverge from the rules, for example if the targets were written as "CtvProstate" and "PtvProstate", then each target volume for the entire patient were sometimes compromised. Although rare, this led mistakes to occur in clusters. However, even when GPT-4 was incorrect, it was typically incorrect in a consistent manner, which could allow for possible algorithmic corrections. Finally, hallucinations were essentially non-existent in this application, possibly due to the lack of freedom afforded to GPT-4 in its responses. Although mis-classifications could be regarded as a hallucination, there were none that were more egregious than an error that a human might make.

Some limitations of this study or method are as follows. Errors may be difficult to address depending on the use-case, but may be alleviated by inspection of the application log files. Since the labels are being standardized to a high degree of accuracy, assuming there is a specific set of structure names of interest for a study, structures being mislabeled should be relatively easy to identify since the vast majority will be correct. Some structure names cannot be easily identified with a particular organ, for example, "CTV\_5000". In this case, the desired output might be "CTV\_Prostate\_5000". This issue could be alleviated by considering more information than the structure name alone. As already mentioned, it was difficult to determine which structures should be considered in determining accuracy, especially since this study is the first to consider LLMs for re-labeling, however our intent was mainly to consider structure names that would be of interest in a data-centered study in radiation oncology (see Section \ref{sec:methods} for more details).


Future work related to this concept may include fine-tuning a smaller foundation model with the raw labels and the ground truth labels. In this way, a smaller LLM, perhaps run locally, may be competitive with the larger LLMs. In general, the performance may possibly be enhanced by providing additional information beyond the structure names alone, such as CT information, structure volume sizes, or dose information, where the information is textual. For example, the volume size, minimum dose, and maximum dose to the structure could be used in the prompts in addition to the structure names for each structure. Finally, as foundation models gain multi-modality capabilities, the AAPM TG-263 report may be included alongside visual representations of the structure set, CT, dose, etc.

\section{Conclusion}

Our results demonstrate that foundation LLMs, and specifically GPT-4, are a compelling new tool for re-labeling structure names in real-world clinical scenarios for radiation oncology, although not perfect. As LLMs continue to advance, the benchmark results presented in this work are expected to be exceeded. It is of our opinion that going forward, LLMs will be the ideal method for applying the AAPM TG-263 standard in radiotherapy clinics.

\appendix
\section{Prostate case}
\begin{lstlisting}
## GPT-4 RESPONSE ##
Reasoning:
OtvProstate: This is an optimization target volume for the prostate. According to the guidelines, optimization volumes should be prefixed with a 'z'. Therefore, it should be renamed to zOTV_Prostate.
bladder: This is a standard structure name and is already in the correct format.
fem_head_r: This represents the right femur head. According to the guidelines, it should be renamed to Femur_Head_R.
fem_head_l: This represents the left femur head. According to the guidelines, it should be renamed to Femur_Head_L.
rectum: This is a standard structure name and is already in the correct format.
penile_bulb: This is a standard structure name and is already in the correct format.
bowel_large: This is a standard structure name and is already in the correct format.
CtvProstate: This is a clinical target volume for the prostate. According to the guidelines, it should be renamed to CTV_Prostate.
svprox: This represents the proximal seminal vesicle. According to the guidelines, it should be renamed to SeminalVes_Prox.
BODY: This represents the body. According to the guidelines, it should be renamed to Body.
body-otv: This represents the body with the optimization target volume subtracted out. According to the guidelines, it should be renamed to Body-OTV.
sv: This represents the seminal vesicle. According to the guidelines, it should be renamed to SeminalVes.
bowel_small: This is a standard structure name and is already in the correct format.
rectal tip: This represents the tip of the rectum. According to the guidelines, it should be renamed to Rectal_Tip.
stv: This represents the scanning target volume. According to the guidelines, it should be renamed to STV.
prostate: This is a standard structure name and is already in the correct format.
FD_Artifact: This represents a fiducial artifact. According to the guidelines, it should be renamed to FD_Artifact.
FD_Air: This represents a fiducial air. According to the guidelines, it should be renamed to FD_Air.
LET: This is not a standard structure name and should be left as-is.

Re-labeled structure names:
{'OtvProstate': 'zOTV_Prostate', 'bladder': 'Bladder', 'fem_head_r': 'Femur_Head_R', 'fem_head_l': 'Femur_Head_L', 'rectum': 'Rectum', 'penile_bulb': 'PenileBulb', 'bowel_large': 'Bowel_Large', 'CtvProstate': 'CTV_Prostate', 'svprox': 'SeminalVes_Prox', 'BODY': 'Body', 'body-otv': 'Body-OTV', 'sv': 'SeminalVes', 'bowel_small': 'Bowel_Small', 'rectal tip': 'Rectal_Tip', 'stv': 'STV', 'prostate': 'Prostate', 'FD_Artifact': 'FD_Artifact', 'FD_Air': 'FD_Air', 'LET': 'LET'}

## RE-LABELING ##
OtvProstate	-------->	zOTV_Prostate
bladder	-------->	Bladder
fem_head_r	-------->	Femur_Head_R
fem_head_l	-------->	Femur_Head_L
rectum	-------->	Rectum
penile_bulb	-------->	PenileBulb
bowel_large	-------->	Bowel_Large
0_Fiducials not considered by GPT-4... unchanged
CtvProstate	-------->	CTV_Prostate
svprox	-------->	SeminalVes_Prox
BODY	-------->	Body
body-otv	-------->	Body-OTV
sv	-------->	SeminalVes
bowel_small	-------->	Bowel_Small
rectal tip	-------->	Rectal_Tip
stv	-------->	STV
prostate	-------->	Prostate
FD_Artifact	-------->	FD_Artifact
FD_Air	-------->	FD_Air
LET	-------->	LET
\end{lstlisting}

\section{Head and neck case}
\begin{lstlisting}
## GPT-4 RESPONSE ##
Reasoning:

1. BODY: This is the body of the patient. According to TG-263, it should be labeled as Body.
2. brain_stem: This is the brainstem. According to TG-263, it should be labeled as Brainstem.
3. brain_stem_prv: This is the planning organ at risk volume (PRV) for the brainstem. According to TG-263, it should be labeled as Brainstem_PRV.
4. cochlea_l: This is the left cochlea. According to TG-263, it should be labeled as Cochlea_L.
5. cochlea_r: This is the right cochlea. According to TG-263, it should be labeled as Cochlea_R.
6. constrictor prv: This is the PRV for the constrictor muscle. According to TG-263, it should be labeled as Musc_Constrict_PRV.
7. constrictors_p: This is the constrictor muscle. According to TG-263, it should be labeled as Musc_Constrict_P.
8. cord: This is the spinal cord. According to TG-263, it should be labeled as SpinalCord.
9. cord_prv: This is the PRV for the spinal cord. According to TG-263, it should be labeled as SpinalCord_PRV.
10. crico_p_inlet: This is the cricopharyngeus. According to TG-263, it should be labeled as Cricopharyngeus.
11. esophagus: This is the esophagus. According to TG-263, it should be labeled as Esophagus.
12. esophagus_cerv: This is the cervical part of the esophagus. According to TG-263, it should be labeled as Esophagus_Cerv.
13. ext_aud_canal_l: This is the left external auditory canal. According to TG-263, it should be labeled as ExtAudCanal_L.
14. ext_aud_canal_r: This is the right external auditory canal. According to TG-263, it should be labeled as ExtAudCanal_R.
15. add 63: This is not a standard structure name and does not follow TG-263 guidelines. It is unclear what this structure represents.
16. hot 63: This is not a standard structure name and does not follow TG-263 guidelines. It is unclear what this structure represents.
17. larynx: This is the larynx. According to TG-263, it should be labeled as Larynx.
18. larynx-ctv: This represents the larynx with the CTV subtracted out. According to TG-263, it should be labeled as Larynx-CTV.
19. larynx-ptv: This represents the larynx with the PTV subtracted out. According to TG-263, it should be labeled as Larynx-PTV.
20. lips: These are the lips. According to TG-263, it should be labeled as Lips.
21. mandible: This is the mandible. According to TG-263, it should be labeled as Bone_Mandible.
22. mastoid_l: This is the left mastoid. According to TG-263, it should be labeled as Bone_Mastoid_L.
23. mastoid_r: This is the right mastoid. According to TG-263, it should be labeled as Bone_Mastoid_R.
24. oral_cavity: This is the oral cavity. According to TG-263, it should be labeled as Cavity_Oral.
25. oral_cavity-ptv: This represents the oral cavity with the PTV subtracted out. According to TG-263, it should be labeled as Cavity_Oral-PTV.
26. parotid_l: This is the left parotid gland. According to TG-263, it should be labeled as Parotid_L.
27. parotid_l_prv: This is the PRV for the left parotid gland. According to TG-263, it should be labeled as Parotid_PRV_L.
28. PTV_7000: This is the PTV with a dose of 7000 cGy. According to TG-263, it should be labeled as PTV_7000.
29. PTV_6300: This is the PTV with a dose of 6300 cGy. According to TG-263, it should be labeled as PTV_6300.
30. PTV_5600: This is the PTV with a dose of 5600 cGy. According to TG-263, it should be labeled as PTV_5600.
31. semi_cir_canal_l: This is the left semicircular canal. According to TG-263, it should be labeled as SemicirCanal_L.
32. semi_cir_canal_r: This is the right semicircular canal. According to TG-263, it should be labeled as SemicirCanal_R.
33. sub_mandib_l: This is the left submandibular gland. According to TG-263, it should be labeled as Glnd_Submand_L.
34. thyroid: This is the thyroid gland. According to TG-263, it should be labeled as Glnd_Thyroid.
35. ctv_5600: This is the CTV with a dose of 5600 cGy. According to TG-263, it should be labeled as CTV_5600.
36. ctv_6300: This is the CTV with a dose of 6300 cGy. According to TG-263, it should be labeled as CTV_6300.
37. ctv_7000: This is the CTV with a dose of 7000 cGy. According to TG-263, it should be labeled as CTV_7000.
38. gtv_7000: This is the GTV with a dose of 7000 cGy. According to TG-263, it should be labeled as GTV_7000.
39. oral_cavity-ctv: This represents the oral cavity with the CTV subtracted out. According to TG-263, it should be labeled as Cavity_Oral-CTV.
40. brain: This is the brain. According to TG-263, it should be labeled as Brain.
41. brain avoid: This is an optimization structure for the brain. According to TG-263, it should be labeled as zBrain.
42. nasal_cavity: This is the nasal cavity. According to TG-263, it should be labeled as Cavity_Nasal.
43. brachial_plex_r: This is the right brachial plexus. According to TG-263, it should be labeled as BrachialPlex_R.
44. brachial_plex_l: This is the left brachial plexus. According to TG-263, it should be labeled as BrachialPlex_L.
45. body-ptv: This represents the body with the PTV subtracted out. According to TG-263, it should be labeled as Body-PTV.
46. PTV_70 opti: This is an optimization structure for the PTV with a dose of 70 Gy. According to TG-263, it should be labeled as zPTV_70.
47. EAM Bolus: This is not a standard structure name and does not follow TG-263 guidelines. It is unclear what this structure represents.
48. ANT NECK Bolus: This is not a standard structure name and does not follow TG-263 guidelines. It is unclear what this structure represents.
49. MID NECK Bolus: This is not a standard structure name and does not follow TG-263 guidelines. It is unclear what this structure represents.
50. POST NECK Bolus: This is not a standard structure name and does not follow TG-263 guidelines. It is unclear what this structure represents.

Re-labeled structure names:
{'BODY': 'Body', 'brain_stem': 'Brainstem', 'brain_stem_prv': 'Brainstem_PRV', 'cochlea_l': 'Cochlea_L', 'cochlea_r': 'Cochlea_R', 'constrictor prv': 'Musc_Constrict_PRV', 'constrictors_p': 'Musc_Constrict_P', 'cord': 'SpinalCord', 'cord_prv': 'SpinalCord_PRV', 'crico_p_inlet': 'Cricopharyngeus', 'esophagus': 'Esophagus', 'esophagus_cerv': 'Esophagus_Cerv', 'ext_aud_canal_l': 'ExtAudCanal_L', 'ext_aud_canal_r': 'ExtAudCanal_R', 'add 63': 'add 63', 'hot 63': 'hot 63', 'larynx': 'Larynx', 'larynx-ctv': 'Larynx-CTV', 'larynx-ptv': 'Larynx-PTV', 'lips': 'Lips', 'mandible': 'Bone_Mandible', 'mastoid_l': 'Bone_Mastoid_L', 'mastoid_r': 'Bone_Mastoid_R', 'oral_cavity': 'Cavity_Oral', 'oral_cavity-ptv': 'Cavity_Oral-PTV', 'parotid_l': 'Parotid_L', 'parotid_l_prv': 'Parotid_PRV_L', 'PTV_7000': 'PTV_7000', 'PTV_6300': 'PTV_6300', 'PTV_5600': 'PTV_5600', 'semi_cir_canal_l': 'SemicirCanal_L', 'semi_cir_canal_r': 'SemicirCanal_R', 'sub_mandib_l': 'Glnd_Submand_L', 'thyroid': 'Glnd_Thyroid', 'ctv_5600': 'CTV_5600', 'ctv_6300': 'CTV_6300', 'ctv_7000': 'CTV_7000', 'gtv_7000': 'GTV_7000', 'oral_cavity-ctv': 'Cavity_Oral-CTV', 'brain': 'Brain', 'brain avoid': 'zBrain', 'nasal_cavity': 'Cavity_Nasal', 'brachial_plex_r': 'BrachialPlex_R', 'brachial_plex_l': 'BrachialPlex_L', 'body-ptv': 'Body-PTV', 'PTV_70 opti': 'zPTV_70', 'EAM Bolus': 'EAM Bolus', 'ANT NECK Bolus': 'ANT NECK Bolus', 'MID NECK Bolus': 'MID NECK Bolus', 'POST NECK Bolus': 'POST NECK Bolus'}

## RE-LABELING ##
BODY	-------->	Body
brain_stem	-------->	Brainstem
brain_stem_prv	-------->	Brainstem_PRV
cochlea_l	-------->	Cochlea_L
cochlea_r	-------->	Cochlea_R
constrictor prv	-------->	Musc_Constrict_PRV
constrictors_p	-------->	Musc_Constrict_P
cord	-------->	SpinalCord
cord_prv	-------->	SpinalCord_PRV
crico_p_inlet	-------->	Cricopharyngeus
esophagus	-------->	Esophagus
esophagus_cerv	-------->	Esophagus_Cerv
ext_aud_canal_l	-------->	ExtAudCanal_L
ext_aud_canal_r	-------->	ExtAudCanal_R
add 63	-------->	add 63
hot 63	-------->	hot 63
larynx	-------->	Larynx
larynx-ctv	-------->	Larynx-CTV
larynx-ptv	-------->	Larynx-PTV
lips	-------->	Lips
mandible	-------->	Bone_Mandible
mastoid_l	-------->	Bone_Mastoid_L
mastoid_r	-------->	Bone_Mastoid_R
oral_cavity	-------->	Cavity_Oral
oral_cavity-ptv	-------->	Cavity_Oral-PTV
parotid_l	-------->	Parotid_L
parotid_l_prv	-------->	Parotid_PRV_L
PTV_7000	-------->	PTV_7000
PTV_6300	-------->	PTV_6300
PTV_5600	-------->	PTV_5600
semi_cir_canal_l	-------->	SemicirCanal_L
semi_cir_canal_r	-------->	SemicirCanal_R
sub_mandib_l	-------->	Glnd_Submand_L
thyroid	-------->	Glnd_Thyroid
ctv_5600	-------->	CTV_5600
ctv_6300	-------->	CTV_6300
ctv_7000	-------->	CTV_7000
gtv_7000	-------->	GTV_7000
z_ptv_63 opti not considered by GPT-4... unchanged
z_ptv_56 opti not considered by GPT-4... unchanged
oral_cavity-ctv	-------->	Cavity_Oral-CTV
brain	-------->	Brain
brain avoid	-------->	zBrain
nasal_cavity	-------->	Cavity_Nasal
brachial_plex_r	-------->	BrachialPlex_R
brachial_plex_l	-------->	BrachialPlex_L
body-ptv	-------->	Body-PTV
PTV_70 opti	-------->	zPTV_70
0_PTV_7000 not considered by GPT-4... unchanged
0_PTV_6300 not considered by GPT-4... unchanged
0_PTV_5600 not considered by GPT-4... unchanged
EAM Bolus	-------->	EAM Bolus
ANT NECK Bolus	-------->	ANT NECK Bolus
MID NECK Bolus	-------->	MID NECK Bolus
POST NECK Bolus	-------->	POST NECK Bolus
\end{lstlisting}

\section{Thorax case}
\begin{lstlisting}
Reasoning:

1. BODY: This is the entire body of the patient. According to TG-263, it should be labeled as Body.
2. SCF_r LN PTV: This is the planning target volume (PTV) for the right supraclavicular lymph node (LN). It should be labeled as PTV_LN_Sclav_R.
3. lung_l: This is the left lung. It should be labeled as Lung_L.
4. lung_r: This is the right lung. It should be labeled as Lung_R.
5. Scar wire: This is a fiducial marker placed at the site of a scar. It should be labeled as FD_Scar.
6. med wire: This is a fiducial marker placed at a medial location. It should be labeled as FD_Med.
7. lat wire: This is a fiducial marker placed at a lateral location. It should be labeled as FD_Lat.
8. cord: This is the spinal cord. It should be labeled as SpinalCord.
9. esophagus: This is the esophagus. It should be labeled as Esophagus.
10. heart: This is the heart. It should be labeled as Heart.
11. lung_total: This is both lungs. It should be labeled as Lungs.
12. thyroid: This is the thyroid gland. It should be labeled as Glnd_Thyroid.
13. body-ptv: This is the body with the planning target volume subtracted out. It should be labeled as Body-PTV.
14. CW_r PTV: This is the planning target volume for the right chest wall. It should be labeled as PTV_Chestwall_R.
15. CW_r PTV_EVAL: This is the evaluation planning target volume for the right chest wall. It should be labeled as PTV_Chestwall_R_Eval.
16. trachea: This is the trachea. It should be labeled as Trachea.
17. hum_head_r: This is the right humeral head. It should be labeled as HumeralHead_R.
18. Bolus_0.3cm: This is a bolus of 0.3 cm. It should be labeled as Bolus_03.
19. IDL 50perc CW: This is an isodose line at 50% for the chest wall. It should be labeled as IDL50_Chestwall.
20. surgical clips: These are surgical clips. They should be labeled as Clips_Surgical.
21. CW_r CTV: This is the clinical target volume for the right chest wall. It should be labeled as CTV_Chestwall_R.
22. carina: This is the carina of the trachea. It should be labeled as Carina.
23. liver: This is the liver. It should be labeled as Liver.
24. chestwall_liver: This is the chest wall and liver. It should be labeled as Chestwall_Liver.
25. IDL 50p CW_flash: This is an isodose line at 50% for the chest wall with flash. It should be labeled as IDL50_Chestwall_Flash.
26. SCF_r LN CTV: This is the clinical target volume for the right supraclavicular lymph node. It should be labeled as CTV_LN_Sclav_R.
27. IDL 50perc SC: This is an isodose line at 50% for the spinal cord. It should be labeled as IDL50_SpinalCord.
28. CouchSurface: This is the surface of the couch. It should be labeled as CouchSurface.
29. CouchInterior: This is the interior of the couch. It should be labeled as CouchInterior.

Re-labeled structure names:
{'BODY': 'Body', 'SCF_r LN PTV': 'PTV_LN_Sclav_R', 'lung_l': 'Lung_L', 'lung_r': 'Lung_R', 'Scar wire': 'FD_Scar', 'med wire': 'FD_Med', 'lat wire': 'FD_Lat', 'cord': 'SpinalCord', 'esophagus': 'Esophagus', 'heart': 'Heart', 'lung_total': 'Lungs', 'thyroid': 'Glnd_Thyroid', 'body-ptv': 'Body-PTV', 'CW_r PTV': 'PTV_Chestwall_R', 'CW_r PTV_EVAL': 'PTV_Chestwall_R_Eval', 'trachea': 'Trachea', 'hum_head_r': 'HumeralHead_R', 'Bolus_0.3cm': 'Bolus_03', 'IDL 50perc CW': 'IDL50_Chestwall', 'surgical clips': 'Clips_Surgical', 'CW_r CTV': 'CTV_Chestwall_R', 'carina': 'Carina', 'liver': 'Liver', 'chestwall_liver': 'Chestwall_Liver', 'IDL 50p CW_flash': 'IDL50_Chestwall_Flash', 'SCF_r LN CTV': 'CTV_LN_Sclav_R', 'IDL 50perc SC': 'IDL50_SpinalCord', 'CouchSurface': 'CouchSurface', 'CouchInterior': 'CouchInterior'}

## RE-LABELING ##
BODY	-------->	Body
SCF_r LN PTV	-------->	PTV_LN_Sclav_R
lung_l	-------->	Lung_L
lung_r	-------->	Lung_R
Scar wire	-------->	FD_Scar
med wire	-------->	FD_Med
lat wire	-------->	FD_Lat
cord	-------->	SpinalCord
esophagus	-------->	Esophagus
heart	-------->	Heart
lung_total	-------->	Lungs
thyroid	-------->	Glnd_Thyroid
body-ptv	-------->	Body-PTV
CW_r PTV	-------->	PTV_Chestwall_R
CW_r PTV_EVAL	-------->	PTV_Chestwall_R_Eval
trachea	-------->	Trachea
hum_head_r	-------->	HumeralHead_R
Bolus_0.3cm	-------->	Bolus_03
IDL 50perc CW	-------->	IDL50_Chestwall
surgical clips	-------->	Clips_Surgical
CW_r CTV	-------->	CTV_Chestwall_R
carina	-------->	Carina
x body_con0.3 not considered by GPT-4... unchanged
x Scar exp 2cm not considered by GPT-4... unchanged
liver	-------->	Liver
chestwall_liver	-------->	Chestwall_Liver
x body_exp0.5 not considered by GPT-4... unchanged
x scar ex u body not considered by GPT-4... unchanged
x scar ex skin not considered by GPT-4... unchanged
x Artifact not considered by GPT-4... unchanged
IDL 50p CW_flash	-------->	IDL50_Chestwall_Flash
SCF_r LN CTV	-------->	CTV_LN_Sclav_R
IDL 50perc SC	-------->	IDL50_SpinalCord
CouchSurface	-------->	CouchSurface
CouchInterior	-------->	CouchInterior
\end{lstlisting}

\bibliography{refs}
\bibliographystyle{unsrt}

\end{document}